\documentclass[twocolumn,aps,prb,twocolum,superscriptaddress,bibnotes,amsmath,amssymb,floatfix,footinbib,longbibliography]{revtex4-2}
\usepackage[markup=nocolor, authormarkupposition=left]{changes}
\usepackage{soul}
\usepackage[utf8]{inputenc}
\usepackage[english]{babel}
\usepackage{amsmath,amsfonts,amssymb}
\usepackage[T1]{fontenc}
\usepackage{url}
\usepackage{changes}
\usepackage{soul}
\usepackage{amsmath}
\usepackage{amsfonts}
\usepackage{amssymb}
\usepackage{xcolor}
\usepackage{siunitx}
\usepackage{epstopdf}
\usepackage{graphicx}
\usepackage{changes}
\usepackage{float}
\graphicspath{{./Figures/}}
\usepackage[
bookmarks=true,
colorlinks=true,
linkcolor=blue,
urlcolor=blue,
citecolor=blue,
pdftex,
bookmarks=true,
linktocpage=true, 
hyperindex=true
]{hyperref}

\newcommand{\SiN}[0]{Si$_3$N$_4$~}

\begin{document}

\title{Near ultraviolet photonic integrated lasers based on silicon nitride}

\author{Anat Siddharth}
\affiliation{Laboratory of Photonics and Quantum Measurements, Swiss Federal Institute of Technology Lausanne (EPFL), CH-1015 Lausanne, Switzerland}

\author{Thomas Wunderer}
\affiliation{Palo Alto Research Center, Palo Alto, CA 94304 USA}

\author{Grigory Lihachev}
\affiliation{Laboratory of Photonics and Quantum Measurements, Swiss Federal Institute of Technology Lausanne (EPFL), CH-1015 Lausanne, Switzerland}

\author{Andrey S. Voloshin}
\affiliation{Laboratory of Photonics and Quantum Measurements, Swiss Federal Institute of Technology Lausanne (EPFL), CH-1015 Lausanne, Switzerland}

\author{Camille Haller}
\affiliation{Laboratory of Advanced Semiconductors for Photonics and Electronics, Swiss Federal Institute of Technology Lausanne (EPFL), CH-1015 Lausanne, Switzerland}

\author{Rui Ning Wang}
\affiliation{Laboratory of Photonics and Quantum Measurements, Swiss Federal Institute of Technology Lausanne (EPFL), CH-1015 Lausanne, Switzerland}

\author{Mark Teepe}
\affiliation{Palo Alto Research Center, Palo Alto, CA 94304 USA}

\author{Zhihong Yang}
\affiliation{Palo Alto Research Center, Palo Alto, CA 94304 USA}

\author{Junqiu Liu}
\affiliation{Laboratory of Photonics and Quantum Measurements, Swiss Federal Institute of Technology Lausanne (EPFL), CH-1015 Lausanne, Switzerland}

\author{Johann Riemensberger}
\email[]{johann.riemensberger@epfl.ch}
\affiliation{Laboratory of Photonics and Quantum Measurements, Swiss Federal Institute of Technology Lausanne (EPFL), CH-1015 Lausanne, Switzerland}

\author{Nicolas Grandjean}
\affiliation{Laboratory of Advanced Semiconductors for Photonics and Electronics, Swiss Federal Institute of Technology Lausanne (EPFL), CH-1015 Lausanne, Switzerland}

\author{Noble Johnson}
\affiliation{Palo Alto Research Center, Palo Alto, CA 94304 USA}

\author{Tobias J. Kippenberg}
\email[]{tobias.kippenberg@epfl.ch}
\affiliation{Laboratory of Photonics and Quantum Measurements, Swiss Federal Institute of Technology Lausanne (EPFL), CH-1015 Lausanne, Switzerland}

\maketitle
\noindent\textbf{Low phase noise lasers based on the combination of III-V semiconductors and silicon photonics are well established in the near infrared spectral regime. Recent advances in the development of low-loss silicon nitride based photonic integrated resonators have allowed to outperform bulk external diode and fiber lasers in both phase noise and frequency agility in the 1550~nm-telecommunication window. Here, we demonstrate for the first time a hybrid integrated laser composed of a gallium nitride (GaN) based laser diode and a silicon nitride photonic chip based microresonator operating at record low wavelengths as low as 410~nm in the near ultraviolet wavelength region suitable for addressing atomic transitions of atoms and ions used in atomic clocks, quantum computing, or for underwater LiDAR.
By self-injection locking of the Fabry-P\'erot diode laser to a high Q ($0.4 \times 10^6$) photonic integrated microresonator, we reduce the optical phase noise at 461 nm by a factor greater than 100$\times$, limited by the device quality factor and back-reflection.}

\medskip
Photonic integrated lasers that operate in the visible to ultraviolet (UV) spectral regime featuring narrow emission linewidth and low phase noise are required for the miniaturization of photonic systems.
Applications for such systems range from quantum metrology and sensing \cite{moody2021roadmap} based on laser cooled neutral atoms and ions \cite{niffenegger2020integrated}, precision atomic clocks \cite{ludlow2015optical}, underwater laser range-finding \cite{mullen1995application}, interferometric biophotonics \cite{ghisaidoobe2014intrinsic}, or visible spectroscopy \cite{kalashnikov2016infrared, card1988prediction}.
The wide bandgap group III-Nitride semiconductor material family is ideally suited as active materials platform for next generation integrated photonics covering operation wavelengths in the UV and visible spectral regime.
High power III-N laser sources (i.e. GaN and its alloys) are commercially available today and can be found in various products such as Blu-ray players, solid-state lighting devices or modern car headlamps.
However, for neutral atom and ion based quantum information science and metrology applications, conventional III-N laser diodes cannot meet the requirements in terms of emission linewidth (i.e. phase noise) and longitudinal mode stability (i.e. drift) during operation.
Instead, only external-cavity diode lasers using bulk precision optics and gratings, which are frequency tuned via physical adjustment, have achieved a suitable performance, exhibiting kHz linewidth \cite{chi2019tunable}.
Yet, the bulk nature of these laser systems, along with their weight, restricts applications, in particular, for space-based applications. More compact blue lasers have been demonstrated based on crystalline resonators, yet these are not wafer-scale compatible \cite{donvalkar2018self}.
As a result, the development of single frequency III-N lasers has regained interest.
These include distributed feedback laser configurations for single frequency operation \cite{hofstetter1998room, masui2006cw, kang2017dfb, slight2019recent, holguin2020480}.
In contrast, silicon photonics-based lasers using heterogeneous \cite{spott2017heterogeneous} and hybrid integration \cite{jin2021hertz} have enabled scalable, narrow linewidth \cite{li2021reaching} and tunable lasers \cite{lihachev2021ultralow} that outperform bulk external diode and fiber lasers and are already employed at a commercial level in data center interconnects, typically operating in the 1550~nm telecommunication window \cite{jones2019heterogeneously}.
Yet, silicon's bandgap limits access to a shorter wavelength.
Silicon nitride (Si$_3$N$_4$) is a good material to realize low loss integrated photonic circuits in the visible and ultraviolet wavelength spectral region due to a wide bandgap of 4.9~eV, high refractive index 2.09 at 410~nm, CMOS-compatible fabrication, and established commercial foundry processes.
Moreover, major advances in nano-fabrication methods have enabled ultra-low propagation loss waveguides, reaching 1~dB/m \cite{liu2021high}.
Demonstrations using \SiN platform in the visible so far include blue laser based beam-forming \cite{shin2020chip}, biophotonic probes \cite{pan2021biophotonic}, modulators \cite{liang2021robust} and visible photonic integrated lasers \cite{corato2021widely}.
Laser cooling of atoms and ions (e.g. Ca$^+$ at 397~nm, Yb at 399~nm, Sr$^+$ at 420~nm and Sr at 461~nm) requires laser wavelength close to 400 nm (cf. Figure \ref{Fig:Fig1}f) with typically $\mathcal{O}(\mathrm{mW})$ power levels.
Integrating compact III-N laser gain elements with high-performance \SiN-based photonic circuits for single-frequency laser operation with narrow linewidth at wavelengths close to 400~nm has not yet been attained, and the prospect for tunability and stable operation has not been explored in great detail.
Here, the hybrid integration of an AlGaInN laser gain element coupled to a \SiN photonic integrated circuit (PIC) platform featuring laser intrinsic linewidth of $\sim$1.15~MHz is demonstrated with more than 20~dB frequency noise reduction via laser self-injection locking.
Such narrow linewidth laser sources, which via integration of AlN or PZT piezoelectric actuators can also be made frequency-agile, i.e. enable mode hop free scanning over $\mathcal{O}(10~\mathrm{GHz})$ with actuation bandwidth of $\mathcal{O}(10~\mathrm{MHz})$ as recently demonstrated \cite{lihachev2021ultralow}, are ideal candidates for photonic integrated lasers for manipulating trapped-ion/atomic quantum systems or underwater coherent laser ranging.
\section*{Results}
\begin{figure*}[t!]
\centering
\includegraphics[width=1.9\columnwidth]{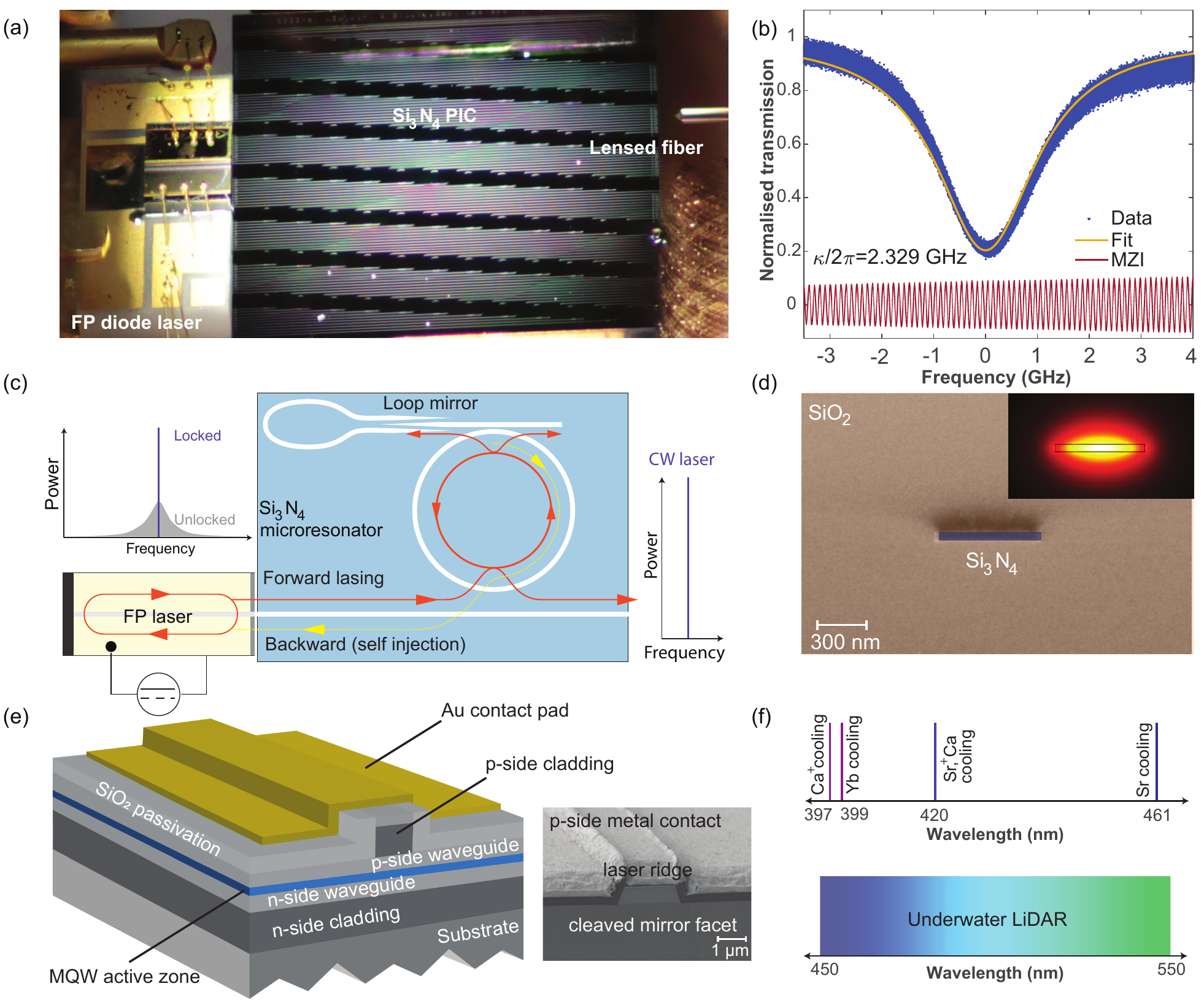}
\caption{
\textbf{Schematic of the hybrid integrated laser system.}
(a)~Photo of the experimental setup showing the Fabry-P{\'e}rot laser diode butt-coupled to the \SiN photonic chip and the output radiation is collected by a lensed fiber.
(b)~\SiN resonance measured using a tunable laser at 461~nm calibrated by a fiber-based Mach-Zehnder interferometer with the free spectral range of 100.12~MHz. The measured loaded linewidth $\kappa/2\pi \approx$ 2.33~GHz corresponding to the loaded quality factor of $0.28\times10^6$. This microresonator did not have a drop-port.
(c)~Principle of laser linewidth narrowing via laser self-injection locking.
(d)~False-colored scanning electron microscope (SEM) image of the sample cross-section, showing a 50 nm thick \SiN buried in SiO$_2$ cladding of total thickness 6~$\mu$m. The inset shows a finite element method (FEM) simulation of the spatial distribution of the TE-mode electric-field amplitude at wavelength 461~nm.
(e)~Schematic showing the laser diode structure with near-UV (410~nm) emission with SEM image of the sample cross-section showing the laser ridge.
(f)~Different applications for III-N integrated lasers presented at wavelength schematics, including underwater LiDAR and ion transitions used for cooling.
}
\label{Fig:Fig1}
\end{figure*}

\noindent \textbf{III-N semiconductor based hybrid self-injection-locked lasers using \SiN integrated photonics.}
Figure \ref{Fig:Fig1} (a, b) depict the experimental setup with a GaN-based Fabry-P{\'e}rot laser diode chip directly butt-coupled to a \SiN photonic chip.
Custom AlGaInN near-UV laser diodes were fabricated on low-defect density native GaN substrates using Metal-Organic Vapor Phase Epitaxy (MOVPE).
The epitaxial growth process was optimized towards high gain and low absorption losses within the laser heterostructure, balancing the optical and electronic performance of the device.
The active zone of the laser diode consists of multiple InGaN quantum wells that are electrically pumped in a pn-junction architecture.
The design was tailored for laser emission near 410~nm and is depicted in Fig.~\ref{Fig:Fig1}e.
In this edge-type emitter configuration, transversal mode confinement is realized through the heterostructure layer stack where the active Multi-Quantum Well (MQW) zone is embedded in (Al)Ga(In)N waveguide and cladding layers featuring different Al-compositions.
Lateral mode confinement is achieved via dry etching a narrow ridge into the p-side of the heterostructure and the creation of optical gain in the QWs right below the etched ridge.
Lateral current confinement and the creation of localized gain is accomplished by ensuring electrical current injection exclusively through an opening in the electrical passivation layer on top of the laser ridge (cf. Figure \ref{Fig:Fig1}e).
The laser ridge had a width of nominally 1.5~$\mu$m and was etched to an optimized depth for laser operation of only one lateral mode.
Laser mirror facets were realized via cleaving along the crystallographic m-plane of the III-N crystal.
The laser resonator cavities were about 1~mm long.
Optical facet coatings were applied to both the front and the rear facet for optimized laser characteristics.
For full continuous wave (CW) operation, individual laser die (LD) were mounted epi-side up onto thermal heat spreader sub-mounts and individually wire-bonded for both the n- and p-side.
The LD-heat spreader sub-mount ensemble was then mounted into a standard TO-5 can.
Special care was taken to allow for excellent laser facet exposure for optimal butt-coupling of the LD to the \SiN chip.
The laser diode showed a characteristic laser threshold of about 70~mA and can produce more than 100~mW of optical output power at a wavelength of 410~nm (cf. Supplementary Figure 1).
The laser is mounted on a thermo-electric cooler for stabilizing its temperature and is operated at 21$^\circ$C.

In addition, in this work we also investigate longer wavelength lasers, notably blue and green laser diodes (LDs) which were provided by Exalos AG \cite{feltin2009broadband}. The emission wavelength of the commercially available blue LD is between 457-464~nm with a threshold current around 18~mA. The green LD has an emission wavelength between 517-523~nm with a threshold current around 38~mA.

The \SiN waveguides and microresonators are fabricated using a subtractive process \cite{luke2013overcoming} and have a uniform height and width of 50~nm and 600~nm, respectively.
Stochiometric \SiN thin films are grown with low-pressure chemical vapor deposition (LPCVD) and etched in fluorine chemistry.
The waveguides and microresonators are defined by deep-ultraviolet stepper (248 nm) lithography.
The cross-section of the microresonators of the \SiN waveguide is depicted in Figure \ref{Fig:Fig1}d.
The waveguides are fully buried in SiO$_2$ cladding of 7~$\mu$m thickness.
The bottom cladding is made of a thermal oxide of 4~$\mu$m thickness, whereas the top cladding is composed of 1~$\mu$m TEOS and 2~$\mu$m low temperature oxide (LTO).
The entire device sits on a 230~$\mu$m thick Si substrate.
The radius of the microresonator is 200~$\mu$m, corresponding to the free spectral range of approximately 107.08~GHz.
The thin \SiN supports the fundamental transverse electric (TE) mode at violet and blue wavelengths as shown in the inset with $\sim$~14.5\% of E-field confined in \SiN core as shown in the inset of Figure \ref{Fig:Fig1}d.
Development of a low-loss photonic platform in the blue and near ultraviolet spectral regime is challenging due to scattering and absorption losses \cite{sorace2018multi}. Rayleigh scattering scales with $\lambda^{-4}$ and material loss also increases as wavelengths approach the materials' bandgap.
Figure~\ref{Fig:Fig1}b presents a cavity linewidth measurement of \SiN microresonator without a drop-port, carried out at 461~nm (see Supplementary Figure 2 for resonance characterization at higher wavelength), which reveals a loaded cavity linewidth $\kappa/2\pi$ = 2.33~GHz and an intrinsic linewidth $\kappa_0/2\pi$ = 1.69~GHz, corresponding to an intrinsic quality factor of $\sim 0.4\times10^6$ (corresponding to a propagation loss of $\sim3$ dB/cm).
The quality factor can be enhanced by designing waveguides with higher aspect ratios and making a thinner \SiN core \cite{morin2021cmos}.
Such a design takes advantage of the lower material loss of the silica cladding and reduces the mode overlap with the sidewalls that minimizes sidewall scattering, which is the primary contributor to loss in high-index-contrast planar waveguides.
The \SiN waveguide has horn-tapered waveguide at the input facet to enhance butt-coupling efficiency with the laser diode and inverse tapered waveguide at the output facet to couple light to the lensed fiber.
The butt-coupling scheme gives an overall insertion loss of ${\sim 7.5 ~{\rm dB}}$ (diode-PIC-lensed fiber), which we measure by comparing the free-space output power of the laser with the fiber-coupled power in the free-running laser regime.
The insertion loss can be further reduced by using bi-layer silicon nitride edge couplers \cite{lin2021low} that enhance fiber-to-chip light coupling in the visible wavelength regime.

\begin{figure*}[t!]
\centering
\includegraphics[width=1.95\columnwidth]{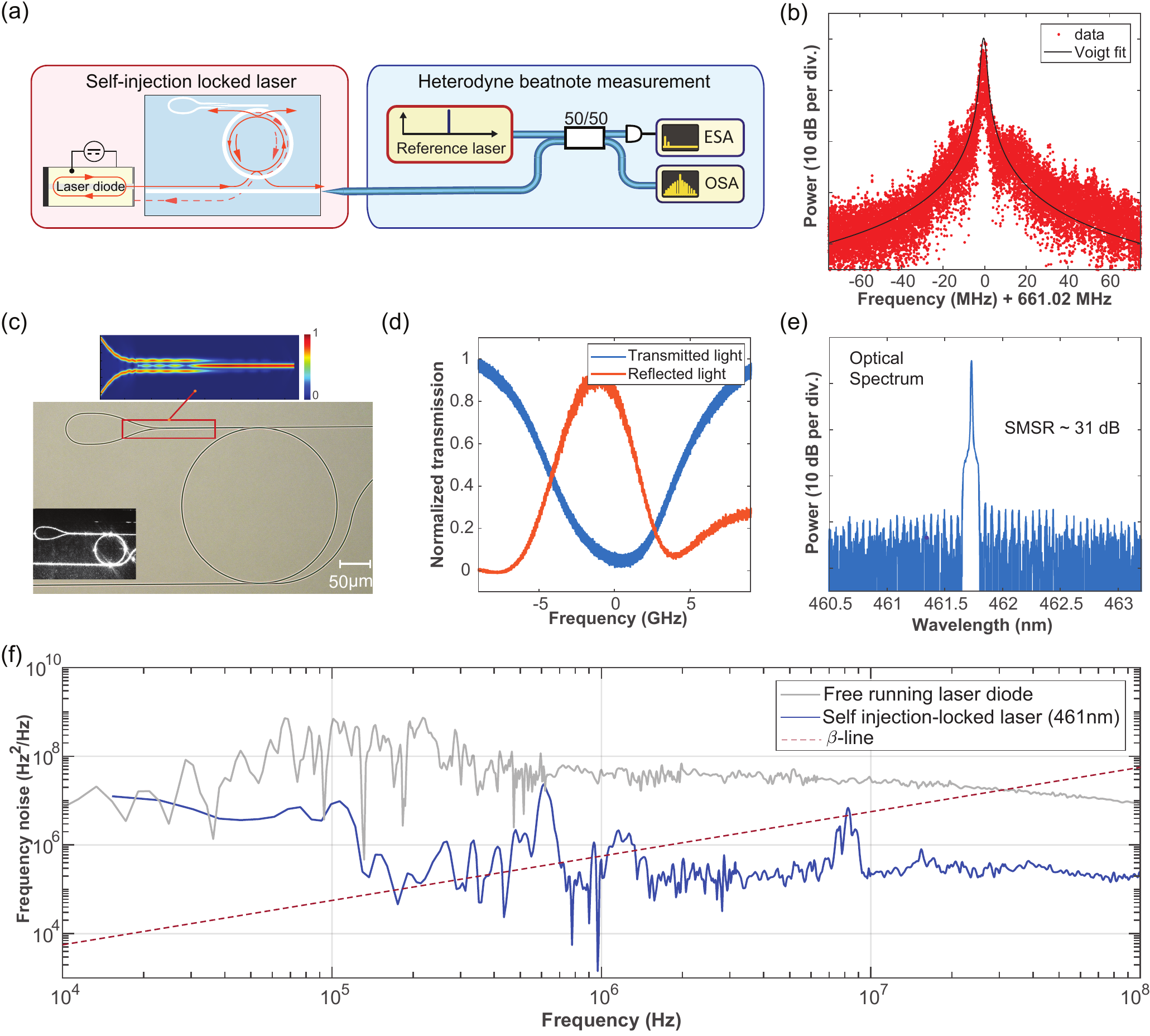}
\caption{
\textbf{III-N hybrid integrated laser performance characterization at 461 nm.}
(a) Experimental scheme of laser frequency noise measurement using heterodyne beat with the reference laser (Toptica DLC DL pro HP) at 461 nm.
(b) Heterodyne beat signal between the injection-locked laser and the reference laser. The measured beat signal is fitted with Voigt profile with Lorentzian FWHM $\sim$1.156~MHz and Gaussian FWHM $\sim$1.618~MHz.
(c) Microscopic image of the 107.08 GHz microresonator with loop-mirror as reflector at the drop-port. The input to the loop-mirror is a adiabatic tapered splitter comprising of three tapered waveguides as shown in the inset, depicting FDTD simulation of the adiabatic tapered splitter. Another inset shows a photograph of the device with 461 nm wavelength light coupled into it.
(d) Optical transmission and reflection spectra of the device shown in (c), measured using the reference laser.
(e) Optical spectrum of the self-injection locked Fabry-P\'erot laser showing emission at 461.5~nm with 31~dB side-mode suppression ratio (SMSRs).
(f) Single sideband frequency noise PSD of the hybrid integrated laser upon self-injection locking to microresonator with FSR 107.08 GHz. The grey line shows the frequency noise of a free-running Fabry-P\'erot laser diode (cf. Supplementary Figure 5). $\beta$-line is shown as a reference (dashed line).
}
\label{Fig:Fig2}
\end{figure*}

Figure~\ref{Fig:Fig1}c illustrates the laser self-injection locking principle.
The key parameters that influence self-injection locking are the quality factor of the microresonator, Rayleigh backscattering and optical feedback phase (optical phase of backscattered light).
We tune the current of the laser diode and thus sweep the relative frequency between the laser and the resonator modes to attain self-injection locking.
When the frequency of the light emitted from the laser diode is close to a high-${Q}$ resonance of the Si$_3$N$_4$ microresonator, laser self-injection locking takes place (see Supplementary Figure 3 for cavity transmission trace at different operating points).
The process occurs due to the coupling of counter-propagating microresonator modes induced by Rayleigh scattering \cite{vassiliev1997injection, raja2019electrically} and the light reflected by the loop-mirror present at the drop-port of the microresonator as shown in Figure \ref{Fig:Fig2}c. Figure \ref{Fig:Fig2}d shows the optical transmission and reflection spectra of a critically coupled device having such a loop mirror as a reflector at its drop-port \cite{wang2016ultra}. This configuration provides a frequency-selective narrowband optical feedback to the laser, leading to a single-frequency operation and a reduction in the laser's frequency noise within the locking range \cite{savchenkov2019self}. The self-injection locked laser will not hop within the locking range. Still, it will hop if we use a different locking state which depends on the current applied on the laser diode, its temperature and the feedback phase from the microresonator (see Supplementary Figure 4 for locking range of the self-injection locked laser).

\noindent \textbf{Laser frequency noise measurements.} Figure \ref{Fig:Fig2}a illustrates the experimental scheme to measure the frequency noise of the laser. 
The laser frequency noise is measured by performing heterodyne beat-note spectroscopy with a tunable external cavity diode laser (Toptica DLC DL pro HP) with a central wavelength at 461~nm as the reference. 
The electrical output of the photodiode is fed to a spectrum analyzer (Rhode \& Schwarz FSW43).
Figure \ref{Fig:Fig2}b shows the heterodyne beat-note of the self-injection locked laser with the reference laser.
The spectrum is fitted with the Voigt profile which provides information about the Lorentzian and Gaussian contribution to the frequency noise of the laser.
The Lorentzian part is linked to the white noise and defines the intrinsic linewidth, whereas the Gaussian part corresponds to the 1/f (flicker) and technical noise of the laser \cite{stephan2005laser}.
From fitting, we extract a Lorentzian linewidth of 1.156~MHz and a Gaussian linewidth of 1.618~MHz. The frequency noise measurement is limited by the frequency noise of the reference laser, and its linewidth is 500~kHz according to its specifications.

Figure \ref{Fig:Fig2}f shows the frequency noise spectra of the free running Fabry Pe\'rot laser and the self-injection locked laser.
The frequency noise of the laser is determined via Welch’s method \cite{welch1967use} from a time sampling trace of the in-phase and quadrature components of the beat-note.
The single sided phase noise power spectral density (PSD) $S_{\phi}(f)$ was converted to frequency noise $S_{\nu}(f)$ according to $S_{\nu}(f)= f^2\cdot S_{\phi}(f)$. The self-injection locked laser optical spectrum is shown in Figure \ref{Fig:Fig2}e indicating laser emission wavelength at 461.5 nm with side-mode suppression ratio of 31~dB.
We also use the beta-line to quantify the linewidth of the self-injection locked laser by integrating the PSD from the intersection of the frequency noise curve with the beta-line $S_{\nu}(f)=8\rm{ln}(2)\rm{f} /\pi^2$ down to the integration time of the measurement.
The integrated frequency noise $\rm{A}$ is used to evaluate the full-width half-maximum measure of the linewidth using $\rm{FWHM}=\sqrt{8\cdot\rm{ln}(2)\cdot\rm{A}}$ \cite{di2010simple}.
We evaluate the FWHM as 3.15~MHz at 10 $\mu$s integration time and 3.55~MHz at 0.1~ms integration time, in agreement with the Voigt fit.
The laser self-injection locking suppresses the frequency noise by at least 20~dB across all frequency offsets.
\newline
\begin{figure}[t!]
\centering
\includegraphics[width=0.5\textwidth]{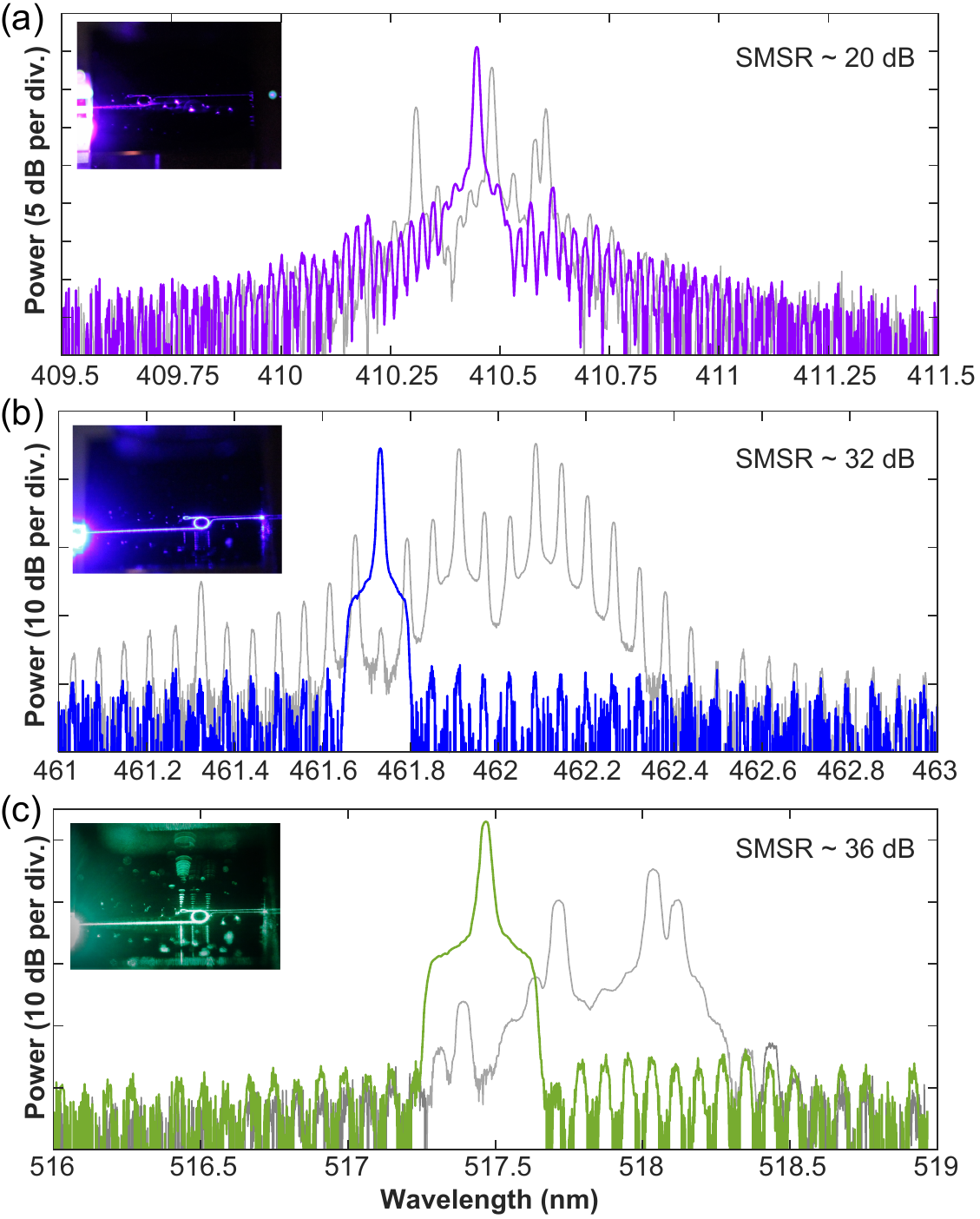}
\caption{ \textbf{Single-frequency lasing in near ultraviolet and visible range.}
Optical spectra of self-injection locked laser states (single frequency) and multi-frequency free-running laser states (grey lines) at different wavelengths ((a) near ultraviolet, (b) blue and (c) green) using the same \SiN photonic chip (D97F4C9). Insets show photos of the self-injection locked lasers at respective wavelengths.}
\label{Fig:Fig3}
\end{figure}

\noindent \textbf{Single-frequency lasing in near ultraviolet and visible regime.} Next, we demonstrate self-injection locked laser in the near ultraviolet (410~nm), as well as in the green wavelength range using the same \SiN photonic chip as shown in Figure \ref{Fig:Fig3}.
Coupling the custom AlGaInN near-UV laser diode with emission at 410~nm (nominal output power of 3.5 mW), we achieve laser self-injection locking at a diode current of 110~mA and voltage of 11 V with a fiber-coupled output power of 0.185~mW.
As shown in Figure \ref{Fig:Fig3}, we achieve single-frequency lasing at 410.3~nm with a side-mode suppression ratio greater than 20~dB.
This constitutes the shortest wavelength hybrid integrated laser based on \SiN.
The inset shows a photograph of the experimental setup where the laser diode is butt-coupled to the \SiN photonic chip.
We clearly observe the scattering of near ultraviolet light around the circumference of the \SiN microresonator, which is indicative of the light travelling inside the cavity.
We also show operation in the visible.
We attained single-frequency lasing at blue (461.8~nm) and green wavelengths (518.6~nm) via self-injection locking.
The fiber-coupled output powers were 1.1~mW and 1.9~mW with side-mode suppression ratios of 32~dB and 36~dB respectively. The blue laser diode was operated at 52~mA and 4.6~V whereas the green laser diode was operated at 83~mA and 6.5~V. Optical spectrum analyzer (Yokogawa AQ6373B) wavelength resolution was set to 0.01 nm.

\section*{Conclusion}
In conclusion, we have demonstrated a hybrid photonic integrated laser operating at near ultraviolet wavelengths as low as 410~nm for the first time. The custom AlGaInN-based laser gain elements were butt-coupled to a \SiN integrated photonic microresonator with high-Q for optical feedback and mode selection.
For the photonic chip, we use \SiN photonic integrated circuits with a thickness of 50~nm and a uniform width of 600~nm delivering an intrinsic quality factor of $0.4\times10^6$ at a wavelength of 461.8~nm.
The high quality factor of our platform ensures single longitudinal mode lasing at green to near-ultraviolet wavelengths with linewidths as low as $\sim$1.15~MHz at 461.8~nm that are traditionally only achieved in bulk external cavity diode lasers.
Further improvements of the device quality factor are possible by decreasing the \SiN waveguide core thickness, improving the waveguide roughness by using the photonic Damascene reflow process \cite{pfeiffer2018ultra, liu2021high}, and by improving the top oxide cladding \cite{bauters2011planar}, as well as by weakening confinement, which however comes at the cost of increasing the minimum bending ring radius and device footprint.
We believe that by optimizing both the coupling between the laser-to-chip and chip-to-fiber and modest improvements of device quality factor, we can achieve multi-mW output powers and sub-100~kHz optical linewidths in the near-ultraviolet region, which would make our systems promising candidates for compact laser implementations for e.g. Sr$^+$ and Ca$^+$ atomic clocks.
\medskip

\begin{footnotesize}

\noindent \textbf{Supplementary Material}: Figures showing the power levels of the custom AlGaIn laser, resonance characterization of the \SiN device at higher wavelength, transmission trace of different self-injection locked states, locking range and stability of self-injection locked laser and spectrum of the free-running Fabry-P\'erot laser can be found in the supplementary material.
 
\noindent \textbf{Funding Information and Disclaimer}: This material is based on research sponsored by Army Research Laboratory under agreement number W911NF-19-2-0345, by the Army Research Office under Cooperative Agreement number W911NF-20-2-0214, and by the Air Force Office of Scientific Research under award number FA9550-21-1-0063. This work was further supported by the Swiss National Science Foundation under grant agreement No. 176563 (BRIDGE). A.S. acknowledges support from the European Space Technology Centre with ESA Contract No. 4000135357/21/NL/GLC/my and J.R. acknowledges support from the Swiss National Science Foundation under grant no. 201923 (Ambizione). The U.S. Government is authorized to reproduce and distribute reprints for Government purposes notwithstanding any copyright notation thereon.

The views and conclusions contained herein are those of the authors and should not be interpreted as necessarily representing the official policies or endorsements, either expressed or implied, of Army Research Laboratory (ARL) or the U.S. Government.

\noindent \textbf{Acknowledgments}:
The authors want to express their gratitude to Christopher Chua and Max Batres at PARC for their contributions in laser fabrication and characterization. We would also like to thank Arslan S. Raja, Jijun He and Viacheslav Snigirev for their assistance in experiments.

\noindent \textbf{Data Availability Statement}: The code and data used to produce the plots within this work will be released on the repository \texttt{Zenodo} upon publication of this preprint.

\noindent\textbf{Correspondence and requests for materials} should be addressed to T.J.K.
\end{footnotesize}
\bibliography{references}

\end{document}